\documentclass[final,3p,times,twocolumn]{elsarticle}
\usepackage{graphicx}
\usepackage{amssymb}
\usepackage{a msmath}
\usepackage{hyperref}
\usepackage{natbib}
\biboptions{sort&compress}
\journal{Physics Letters A}
\usepackage[usenames]{color}
\usepackage[normalem]{ulem}  
\begin{document}                                               
\begin{frontmatter}
\title{
Mass-imbalanced Bose-Einstein condensed mixtures in rotating perturbed trap
}
\author[NZ,ift]{R. Kishor Kumar}
\author[usp]{A. Gammal}
\author[ift]{Lauro Tomio}
\ead{lauro.tomio@unesp.br }
\address[NZ]{Department of Physics, Centre for Quantum Science, and Dodd-Walls Centre for
Photonic and Quantum Technologies, University of Otago, Dunedin 9054, New Zealand.}  
\address[ift]{Instituto de F\'{\i}sica Te\'orica, Universidade Estadual Paulista (UNESP), 01140-070, S\~ao Paulo, Brazil.} 
\address[usp]{Instituto de F\'{i}sica, Universidade de S\~{a}o Paulo, 05508-090, S\~{a}o Paulo, Brazil.}

\begin{abstract}
We consider the mass-imbalanced sensibility for the emergence of vortex patterns in the Bose-Einstein 
condensed binary mixture of rubidium-cesium ($^{85}$Rb-$^{133}$Cs), confined in quasi-two-dimensional 
harmonic traps, with one species linearly perturbed in one direction.  
Non-dipolar coupled species are chosen to highlight mass symmetry effects. We first analyze 
the condensed mixture in the unperturbed non-rotating regime, where 
radial phase separation is verified in the immiscible regime, which occurs for large ratio 
between inter- and intra-species repulsive interactions.  
By going to the linear perturbed regime, the radial 
phase separation that occurs in the immiscible condition splits up with the two densities 
having their maxima at distinct positions. In the rotating regime of both unperturbed and perturbed cases, 
the minimum rotation is determined in terms of the inter-species interaction to observe vortex structures. 
In the immiscible regime a dramatic spatial interchange between the species is verified by increasing the rotation. 
\end{abstract}

\end{frontmatter}

\section{Introduction}\label{sec1}
The first reported experimental realization of a Bose-Einstein condensate (BEC) mixture with two hyperfine spin states of $^{87}$Rb~\cite{myatt1997}, and the corresponding experimental observation of vortices in this two-component hyperfine mixture~\cite{matthews1999},  have motivated several other theoretical and experimental studies with BEC binary mixtures. 
Actually, following these works, it is been quite well established the observation and several studies with binary BEC mixtures 
having different hyperfine states, different atomic isotopes, as well as different atom-atom or atom-molecular 
mixtures~\cite{iso-rbyb,2011Suga,rbk1,2008Tha,2010brtka,rbcs1,rbcs2,rbsr,rbk2,2015Liu,2016lee,erdy,2015polo,2019kuopanportti}.
Further, coreless vortices are also observed in the three-component order parameter with $F=1$, for sodium 
condensates~\cite{coreless}.  More related studies on vortices and vortex lattices observations, with particular focus 
on atomic systems having magnetic dipole moments, in which properties of quantum ferrofluids emerge,
can also be found in the recent review by Martin et al.~\cite{Parker}, where they discuss analytic treatments based on 
the Thomas-Fermi and variational approaches, as well as full numerical simulations. As verified, even with
single atomic species, the studies on generating vortices in dipolar condensates 
lead to the appearance of a rich variety of vortex lattice structures~\cite{2019Bijnen}. 
In view of the actual experimental progress with ultracold dipolar atomic systems~\cite{erdy},
one of the interesting aspects been considered on the properties of rotating dipolar mixtures,
is the possibility for tuning the corresponding inter-species dipolar interactions~\cite{2019-Kishor}.
The studies on vortex patterns in multicomponent BECs are also interesting due to different miscibility properties.
Peculiar vortex structures can be verified in addition to the fundamental Abrikosov's triangular 
lattice~\cite{2009fetter}, such as squared, striped, with domain walls, and rotating 
droplets~\cite{Mueller,vortex-phase,vort-sheet,2c-vort,dbec-vort}. 
When considering ring-trap geometry, the miscibility and stability requirements for binary atomic BECs with 
repulsive interactions was recently studied in Ref.~\cite{2019Chen}.
More recently, within the investigations with binary mixtures of Bose-Einstein condensates, we noticed studies with 
massive core vortices~\cite{2020richaud}, as well as the dynamics of vortex-bright-soliton spontaneous generation 
with small mass imbalance between the species~\cite{2020mukherjee}.

The experiment with the $^{85}$Rb-$^{133}$Cs mixture, reported of having three distinct density 
distributions, depending on the relative atom numbers in each component of BECs~\cite{rbcs1}, was further 
analyzed by using the mean-field theoretical model~\cite{Pattinson2013}. From this analysis, it was observed that 
weak perturbations are provided by the tilt in the magnetic dipole trap, due to larger inelastic three-body losses, which
affect the equilibrium density distributions, displaying different miscibility or phase separations. 
The tilt in the magnetic dipole trap, having small differences between the species, was applied in one of the transverse 
direction, implying in a corresponding shift in the relative trap centers. 
Therefore, by applying a linear-trap perturbation, in addition to the harmonic trap  in a cigar shaped trap,  
one can analyze the effect of the tilt in the magnetic dipole trap and corresponding density distributions.
Further, one could also consider a linear-trap perturbation in two-dimensional coupled systems,
to investigate the effect on the vortex structure patterns.  

Motivated by a previous investigation with dipolar mixtures~\cite{2019-Kishor}, in which a  remarkable 
mass-imbalance sensibility was verified in the miscibility and vortex-pattern distribution of 
symmetric- and asymmetric-dipolar mixtures, 
the focus of the present paper is to study the mass-imbalance effect in the density phase separation 
and corresponding vortex pattern structure, by considering two non-dipolar species in which
the mass symmetry effect can be more clearly evidenced by varying the associated two-body interactions.
For that, in our study we select  the non-dipolar coupled mixture $^{85}$Rb-$^{133}$Cs due to the actual 
interest and experimental possibilities. 
With non-dipolar binary system, the mass-imbalance sensibility can be highlighted
with simple non-perturbed and perturbed pancake-like traps.
We should observe that, among the non-dipolar binary mixtures being studied in cold-atom laboratories,
one can consider atomic elements with small mass difference, as two isotopes of rubidium~\cite{myatt1997} or 
two isotopes of cesium~\cite{weber2003} or with large mass difference, as the binary mixture 
$^{85}$Rb-$^{133}$Cs~\cite{rbcs1,rbcs2}. 
The first case, use to be investigated by model approximations where both species have
identical masses~\cite{2019-adhikari}.  However, in view of the observed mass-imbalance sensibility~\cite{2019-Kishor}, 
we found appropriate to work with the second case where the mass difference cannot be neglected, 
such that relevant related properties on the miscibility and vortex structures can be pointed out. 
As verified, the more mass-symmetric BEC mixtures, as $^{87}$Rb-$^{85}$Rb or the dipolar one $^{164}$Dy-$^{162}$Dy,
can show  triangular,  squared, striped, domain wall, and rotating droplets vortex-lattice structures regarding 
to the ratio between inter- and intra-species contact interaction. 
On the other hand, highly mass-imbalanced mixtures, such as $^{85}$Rb-$^{133}$Cs, are more likely to 
present radial vortex-lattice patterns in the quasi-2D pancake-like configuration. 

Our analysis is performed for a pancake-shaped trap configuration, 
with the trap frequencies $\omega_x=\omega_y\equiv\omega_\rho\ll \omega_z$ (aspect ratio 
$\lambda\equiv\omega_z/\omega_\rho\gg 1$), in which the underlying three-dimensional (3D) system can be 
reduced to a two-dimensional (2D) one. 
The effect of a perturbation in the original pancake-like harmonic trap is verified by adding a 
linear shift of one of the species along the $x$-direction.
In our study, we assume fixed and repulsive the intra-species interactions, $a_{11}=a_{22}$, varying the
inter-species one, $a_{12}$, from attractive to repulsive interactions, which will correspond in going from
miscible to immiscible regimes of the mixture. 
The critical rotation frequencies to generate vortices in the coupled system are verified by introducing 
rotation in the trapped system.  We have also verified the density distribution of the two species and how 
the dynamics of vortices is being modified by increasing the rotation. Guided by a previous 
analysis~\cite{2019-Kishor}, we understand that highly mass-imbalanced mixtures in non-perturbed
harmonic trap and without rotation 
should present radially phase-separated distributions in the immiscible regime where the inter-species 
$a_{12}$ is positive and larger than the repulsive intra-species ones, $a_{ii}$ ($i=1$ and $2$). 
After verifying the effect of the shift-perturbation in one of the trapped species (non-rotating case), 
we study the non-perturbed and perturbed rotating mixtures, in order to observe how the radial 
separations and vortex-pattern structure are affected by the mass asymmetry, with particular focus in the 
immiscible regime.
 
In the next Sect.~\ref{sec2}, we have the mean-field approach applied to rotating binary mixtures 
in a pancake-like trap, which includes a discussion on miscibility properties.
In Sect.~\ref{sec3}, after a brief discussion on our numerical procedure, we present our main results.
Finally, in Sect.~\ref{sec4} we have a summary with concluding remarks.

\section{Mean-field model for rotating binary BEC}\label{sec2}

In our approach for the coupled BEC system, the two atomic species $i=1,2$ with masses $m_i$ are  
assumed with the same number of atoms $N_i\equiv N$, confined in strongly pancake-shaped harmonic 
traps with fixed aspect ratios $\lambda\equiv \lambda_i=\omega_{i,z}/\omega_{i,\perp}=10$,
where $\omega_{i,z}$ and $\omega_{i,\perp}$ are, respectively, the longitudinal and transverse trap frequencies
for the species $i$. We further assume the intra-species scattering lengths are identical and fixed for both species,
with $a_{11}=a_{22}=150a_0$ ($a_0$ is the Bohr radius), such that the relative strength is 
controlled by the inter-species interaction $a_{12}$.
The coupled Gross-Pitaevskii (GP) equation is cast in a dimensionless format, with energy and length 
units given, respectively, by $\hbar \omega_{1,\perp}$ and $l_{\perp,1}\equiv \sqrt{\hbar/(m_1\omega_{1,\perp})}$. 
By taking the first species as the reference in our unit system, in the next we have 
$\omega_\perp\equiv\omega_{1,\perp}$ and 
$l_\perp\equiv l_{\perp,1}$. Correspondingly, the space and time variables are 
such that  ${\bf r}\to l_\perp {\bf r}$ and $t\to \tau/\omega_{\perp}$, when going from full-dimension to 
dimensionless quantities.
 Within these units,  for simplicity we first adjust both trap frequencies as 
 ${m_2\omega_{2,\perp}^2}={m_1\omega_{1,\perp}^2}$, such that the dimensionless non-perturbed 3D trap 
 for both species have the same expression given by 
\begin{equation}
V_{i,3D}({\bf r})=\frac{1}{2}(x^{2}+y^{2}+\lambda^2 z^2) \equiv V_0(x,y) + \frac{1}{2}\lambda^2 z^2,
\label{3Dtrap}
\end{equation}
where $V_0(x,y)$ is the 2D non-perturbed harmonic oscillator. By adjusting the trap frequencies as above, we 
have no explicit mass-dependent factor in the trap potential.
Next, we also assume a large value for the aspect ratios $\lambda$, which allows us to reduce the original 
3D formalism to a 2D one, by the usual factorization for the 3D wave function, $\psi_{i}(x,y,\tau)\chi_i(z)$, 
where $\chi_i(z)\equiv \left(\lambda_i/{\pi}\right)^{1/4}e^{-{\lambda_i z^{2}}/{2}}$. 
In this case, the ground-state energy for the harmonic trap in the $z-$direction is a constant factor to be 
added in the total energy. 
It is safe to assume a common mass-independent transversal wave-function for both components, with 
$\lambda_i=\lambda$, as any possible mass dependence can be absorbed by changing the corresponding aspect ratio.
This approach for the reduction to 2D implies that we also need to alter the nonlinear parameters accordingly, as 
the integration on the $z-$direction will bring us a $\lambda-$dependence in the non-linear parameters.  
So, the corresponding coupled 2D GP equation in the rotating frame is given by
\begin{eqnarray}
\mathrm{i}\frac{\partial \psi_{i}  }{\partial \tau }
&=&{\bigg\{}\frac{-m_{1}}{2m_{i}}{\left(\frac{\partial^2}{\partial x^2}+\frac{\partial^2}{\partial y^2}\right)}
+V_i(x,y)- \Omega L_{z}\nonumber\\
&+&\sum_{j=1,2}g_{ij}|\psi_{j} |^{2}{\bigg\}}
\psi_{i}  
\label{2d-2c}, 
\end{eqnarray}
where 
$\psi_{i}\equiv \psi_{i}(x,y,\tau)$ and $\psi_{i}^\prime\equiv \psi_{i}(x^\prime,y^\prime,\tau)$
are the components of the total 2D wave function, normalized to one, 
$\int_{-\infty}^{\infty}dx\,dy\,|\psi _{i}|^{2}=1.$ 
$L_z$ is the angular momentum operator with $\Omega$ the corresponding rotation parameter (in units of  $\omega_{\perp}$), 
which is common for the two components. 
The two-body contact interactions, which are related to the scattering lengths $a_{ij}$ for the species $i,j = 1,2$ (where
$a_{ij}=a_{ji}$) are given by
\begin{eqnarray}
g_{ij}\equiv \sqrt{2\pi\lambda}
\frac{m_1 a_{ij} N_j}{\mu_{ij}l_\perp},
\label{par}
\end{eqnarray}
where $\mu_{ij} \equiv m_im_j/(m_i+m_j)$ is the reduced mass; and 
we assume $a_{11}=a_{22}$, with the same number of atoms for both species, $N_{1}=N_{2}$.
In the next, the length unit will be adjusted to  $l_\perp = 1\mu$m$ \approx 1.89\times 10^4 a_0$, where 
$a_0$ is the Bohr radius, such that $a_{ij}$ can be conveniently given in terms of $a_0$.
$V_{i=1,2}(x,y)$ is the external potential provided by the harmonic trap, which we assume that the component
$i=1$ can be perturbed by the addition of a linear trap along $x$. It can be expressed by
\begin{equation}\label{trap}
V_i(x,y)=\frac{x^{2}+y^{2}}{2} + \nu_i x =  
\frac{(x+\nu_i)^{2}+y^{2}}{2} - \frac{\nu_i^2}{2}
,\end{equation}
where $\nu_{i=1,2}=0$ for the non-perturbed case; and $\nu_1=1$, with $\nu_2=0$ for the perturbed case.
As indicated by the above Eq.~{\ref{trap}}, in the perturbed case the center of the harmonic trap for the  
species $i=1$ is shifted to $x=-1$.

A relevant property to be considered for coupled mixtures is the miscibility of the components.
By following a simplified energetic approach derived in  Ref.~\cite{Chui-1998} for  
homogeneous systems, one can characterize the transition between miscible and immiscible states 
for systems with repulsive two-body interactions by a criterion, which is independent on the condensate 
atom numbers or trap sizes. Considering this criterion, a coupled system enters in an immiscible
regime for $G_{12}^2 > G_{11} G_{22}$, where $G_{ij}$ are given by the ratio between the 
corresponding two-body scattering lengths $a_{ij}$ and reduced masses $\mu_{ij}$, 
given by $G_{ij}\equiv 2\pi\hbar^2 a_{ij}/\mu_{ij}$.
This condition to enter in an immiscible regime, which also corresponds to 
$g_{12}^2 > g_{11} g_{22}$ [using Eq.~(\ref{par})], for $a_{11}=a_{22}>0$, 
can define a threshold parameter $\delta$, given by
\begin{equation}
\delta\equiv \frac{a_{12}}{a_{11}} > \frac{2\sqrt{m_1m_2}}{m_1+m_2},
\label{delta}\end{equation}
where the right-hand-side is the mass-dependent critical value for the miscible-immiscible transition
of homogeneous mixture. In the present case, this critical value is $\delta\simeq 0.975$, which is not far 
from the equal-mass case, such that we can still consider $\delta=1$ as the approximate value for 
the transition. 

When considering general non-homogeneous coupled mixture, the miscibility was further studied in 
Ref.~\cite{2017jpco}, where a convenient parameter was defined for the miscibility, using the overlap
between the densities, which for the present 2D case is given by
\begin{eqnarray}
\eta = \int \vert \psi_1 \vert \vert \psi_2 \vert \ dx dy = 
\int \sqrt{\vert \psi_1 \vert^2 \vert \psi_2 \vert^2} \ dxdy
,\label{eta}\end{eqnarray}
where both  $\psi_1$ and $\psi_2$ are normalized to one.
It implies in $\eta=1$ for the complete overlap between the two densities and zero in the opposite limit. 
Equivalent to the above definition, one can also find some other suggestions, as in Refs.~\cite{2012-Wen,2017-Bandy},
used as criterion to separate the miscible and immiscible phases, based on the overlap of the densities.

\section{Results for non-rotating and rotating binary mixtures in non-perturbed and perturbed traps}\label{sec3}
We start by a brief explanation on our numerical approach to obtain the results for general mass-imbalanced 
binary mixtures.
As one can verify from the formalism given in section~\ref{sec2}, the mass symmetry appears explicitly 
in the kinetic-energy terms (reflected in the second component of the coupled system), with the 
mass-imbalance factor given by $(m_2/m_1-1)$.  By assuming the species $i=2$ as the larger-mass 
component of the mixture, in case of $^{85}$Rb-$^{133}$Cs the mass-imbalanced factor is 0.5647.
For the atom-atom interactions,  we  assume enough large repulsive intra-species scattering lengths 
$a_{11}=a_{22}=150a_0$, remaining  the inter-species one to be explored from attractive to larger 
repulsive values, through the parameter $\delta$, given by (\ref{delta}). 
\begin{figure*}[h]
\begin{center}
\includegraphics[width=0.8\linewidth]{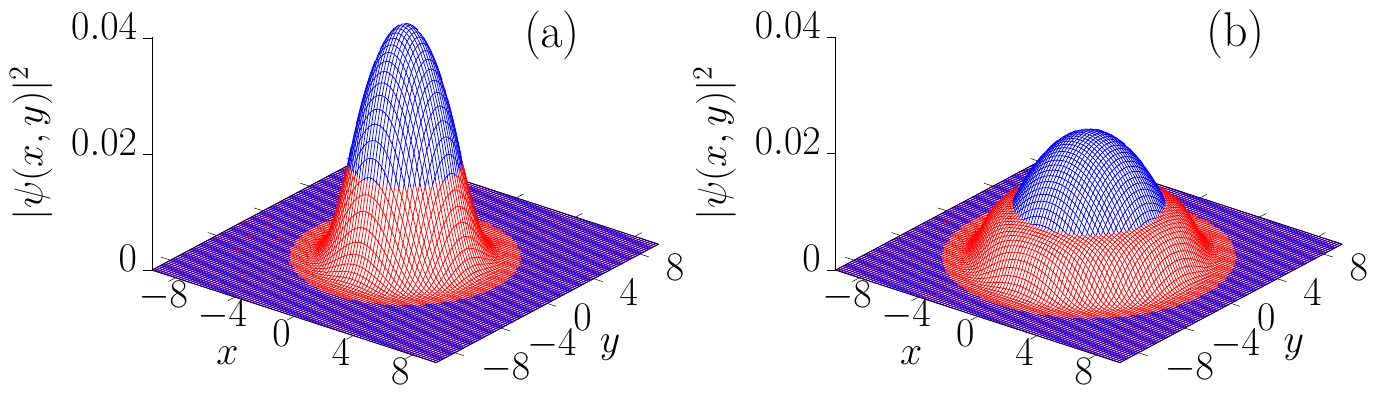}
\includegraphics[width=0.8\linewidth]{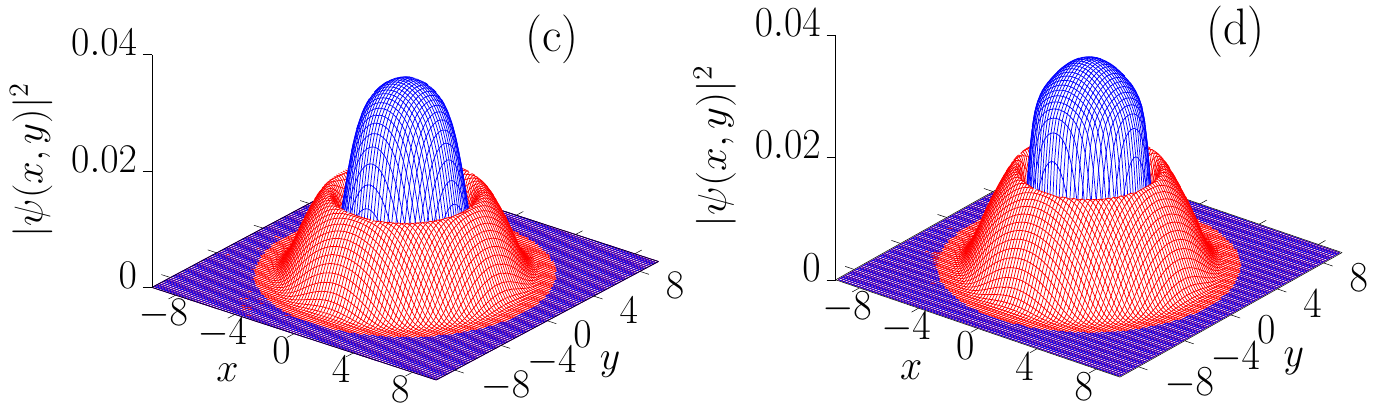}\vspace{0.2cm}
\includegraphics[width=0.8\linewidth]{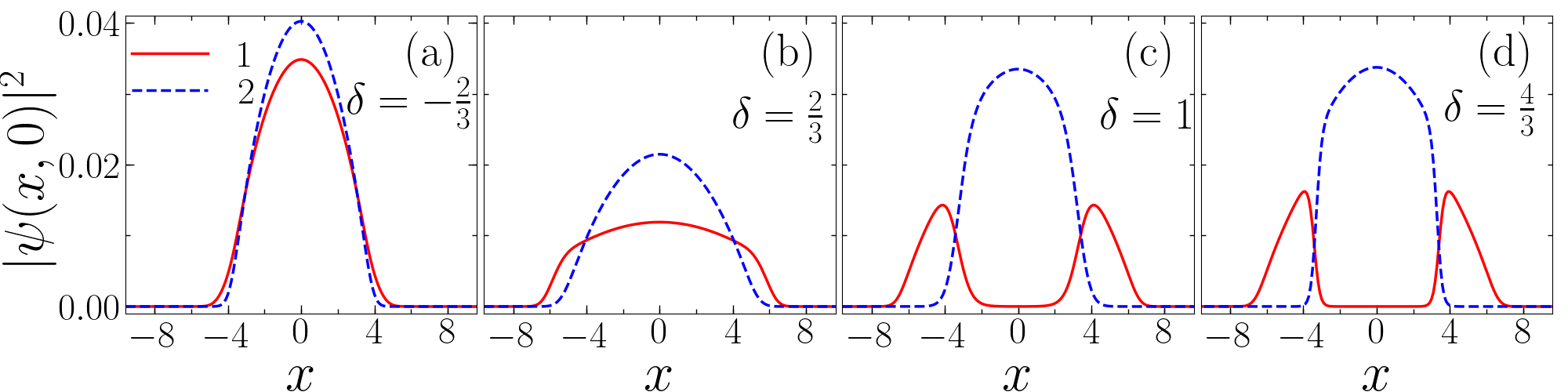}
\end{center}\vspace{-.5cm}
\caption{
The non-perturbed non-rotating ($\Omega=0$) harmonically confined, with aspect ratio $\lambda=10$, 
2D component densities (units of  $l_{\perp}^{-2}$) are shown for the mixture $^{85}$Rb-$^{133}$Cs, 
using surface plots in the upper (a)-(d) panels, in which $a_{12}= -100a_0$ (a), $a_{12}= 100a_0$ (b), 
$a_{12}= 150a_0$ (c), and $a_{12}= 200a_0$ (d).
With $a_{11}=a_{22}=150a_0$ fixed, the corresponding $\delta$ are indicated inside the lower panels, in which
the respective central densities ($y=0$) are presented along the $x-$direction (units of  $l_{\perp}$), 
with the red-solid line referring to $^{85}$Rb (element 1) and the blue-dashed line  to $^{133}$Cs (element 2). 
}\label{f-01}
\end{figure*}
\begin{figure*}[h]
\begin{center}
\includegraphics[width=0.75\linewidth]{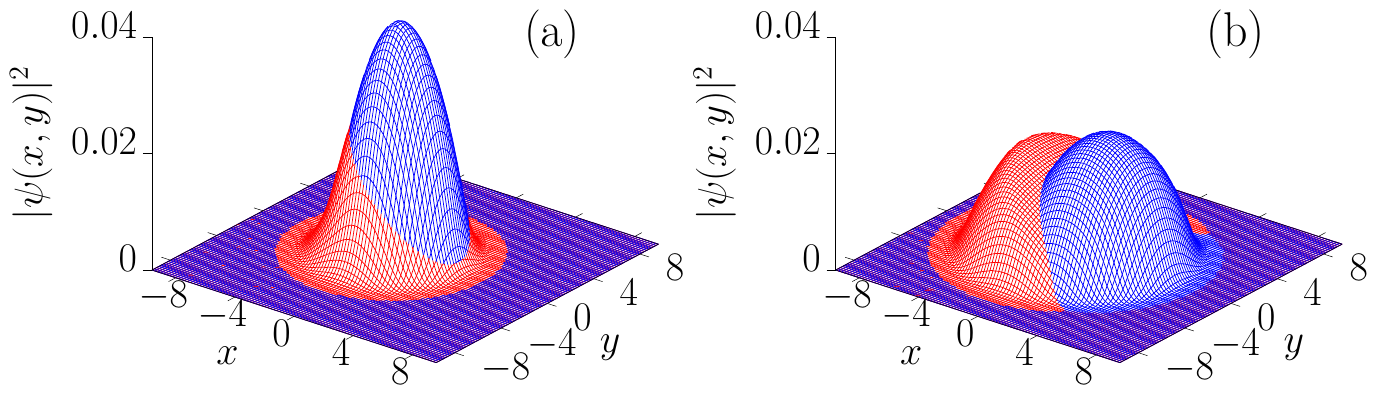}
\includegraphics[width=0.75\linewidth]{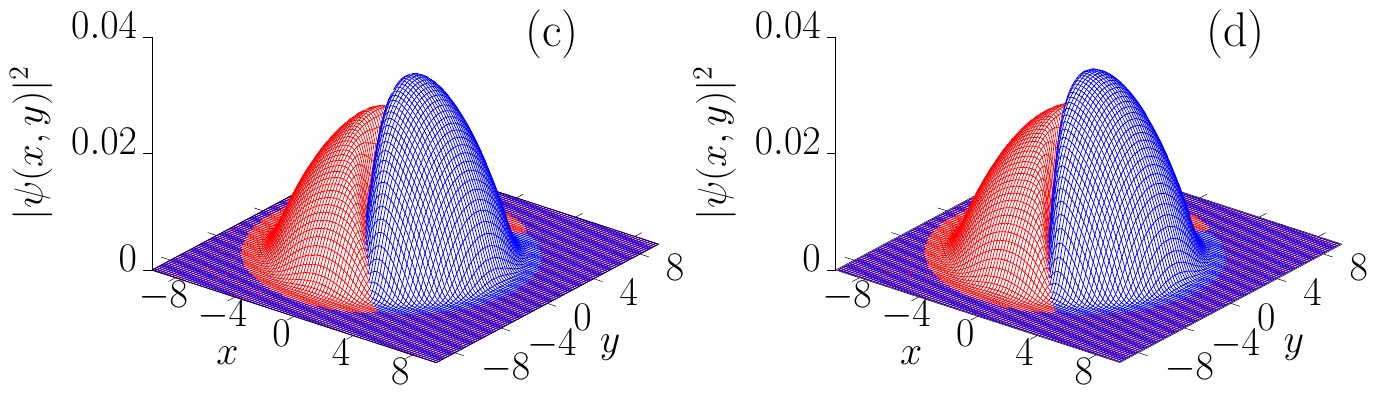}\vspace{0.2cm}
\includegraphics[width=0.75\linewidth]{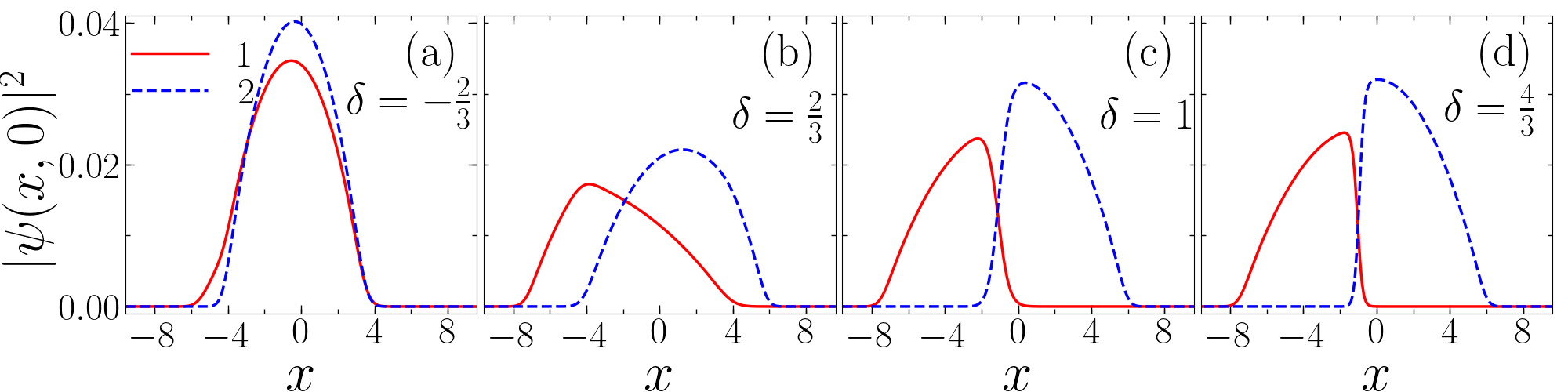}
\includegraphics[width=0.75\linewidth]{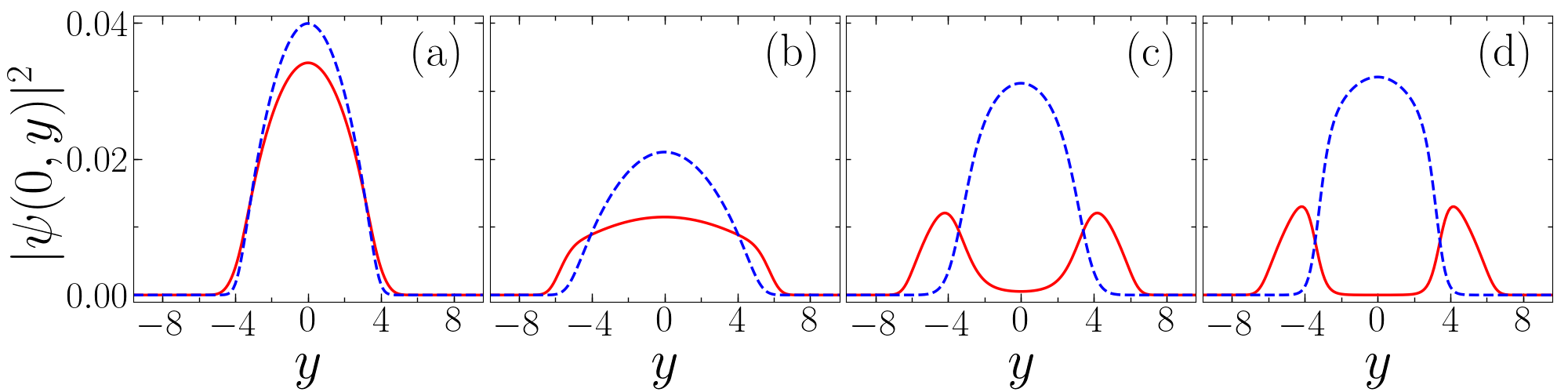}\vspace{-0.2cm}
\caption{
With the harmonic trap linearly perturbed in the $x-$direction ($\nu_1=1, \nu_2=0$), this figure
shows the corresponding panels presented in Fig.~\ref{f-01}. Except for the perturbation, all the
other parameters are the same as used in Fig.~\ref{f-01}. In this case, the two lower row of panels 
are for $|\psi_i(x,0)|^2$ and $|\psi_i(0,y)|^2$, along the $x$ and $y$ directions, respectively.  
}
\label{f-02}
\end{center}
\end{figure*}
The 2D coupled mean-field equation, was solved 
by the split time-step Crank-Nicolson method with an appropriate algorithm discretization as described in   
Ref.~\cite{kk2019}.
As our aim was to study some ground state properties of the mass-imbalanced coupled system, we have 
mainly concerned with time-independent solutions.  As to confirm the stability of our results
in the calculation of non-rotating ground states,  for the initial component wave functions used in the  
time-dependent equation (\ref{2d-2c}),  we assume a simplified Gaussian expression, given by  
$\psi_{i,0}(x,y) = (1/\sqrt{\pi})\exp \left[-(x^2+y^2)/2\right]$. 
However, when studying rotational properties, for the initial component wave-functions we consider
these component wave functions with a single vortex at the center, modulated by a random phase at different 
space points, as  
{\small\begin{equation} \label{initial}
\psi_{i,0}(x,y) = \frac{{(x+\mathrm i y)}}{\sqrt {\pi}}\exp 
\left( -\frac{x^2+y^2}{2}{+2\pi \mathrm i {\cal R}(x,y)}\right),
\end{equation}}
where ${\cal R}(x,y)$ is a random number in the interval $[0,1]$.
This procedure, found to be simpler than the one used in \cite{dbec-vort},
 has been shown to be relevant to address convergence issues, and to
overcome the impact of the initial condition on the final vortex solutions~\cite{kk2019}.
Within our dimensionless variables, for the coupled system we use space step size 0.05 
and time step 0.0005, with the stability being confirmed by performing time evolution
from $t=0$ up to $t=2000$ (units $\omega_\perp^{-1}$).  

In the next sub-sections, our main numerical results are detailed, in which the unperturbed case, which can
be found in several other studies, as already mentioned, is reported to serve as reference to the 
other results we are presenting for the perturbed case, as well as for the rotating cases.

\subsection{Non-rotating, non-perturbed and perturbed system}
First, we present in Fig.~\ref{f-01} the results for the density distributions of the two components within 
non-perturbed harmonic traps, in order to verify the corresponding miscibility properties in non-rotating 
conditions,  $\Omega=0$.  
In the upper (a)-(d) four panels, by using 2D surface plots, we show how the corresponding densities of the two 
elements [$^{85}$Rb and $^{133}$Cs] are spatially distributed, for different inter-species scattering lengths, and 
confined in harmonic traps with the same aspect ratio $\lambda=10$. In the lower (a)-(d) four panels we show the
respective densities [corresponding to the (a)-(d) upper panels] along the $x-$direction, for $y=0$. In all the results,
the intra-species scattering lengths are kept fixed, $a_{11}=a_{22}=150a_0$, with the inter-species fixed at different
values at each panel, such that in the panels (a) we have $a_{12}=-100\,a_0$; in panels (b), $a_{12}=100\,a_0$; 
in panels (c), $a_{12}=150\,a_0$;  and, in panels (c), $a_{12}=200\,a_0$. 
The plots are showing the miscibility to immiscibility transition for non-rotating ($\Omega =0$) binary BEC mixture confined
in harmonic trap with aspect ratio $\lambda=10$. The resulting distribution is such that, as $\delta$ becomes larger than 1, 
the massive component ($i=2$, the $^{133}$Cs) remains with the maximum in the center, with the density of the lighter one 
being reduced to a minimum in the center.
As verified, in the panels (a) of Fig.~\ref{f-01}, the attractive interspecies interaction provides complete 
miscibility (overlap) of the species, with maxima for the densities at the center.
As shown in the panels (c), for $\delta=1$ ($a_{12}=a_{ii}$) we have already almost complete radial phase separations 
of the mixture, which is further characterized in the panels (d), with $a_{12}=200\,a_0$, where $\delta= 4/3$.
These results are consistent with the ones verified in Ref.~\cite{2017-Bandy} for imbalanced-mass binary systems 
in pancake-like non-perturbed harmonic traps.

By considering linearly perturbed trap, our results for the 
miscibility to immiscibility transition of non-rotating ($\Omega =0$) binary BEC are presented in Fig.~\ref{f-02}. 
In the upper panels we have the surface plots with the 2D densities ($|\psi_i(x,y)|^2$) given in the 
$x-y$ plane, whereas in the two lower row of panels we have the corresponding density results given 
for $y=0$ and $x=0$, respectively.  In all these results, we consider four values for the interspecies scattering 
length, such that  $a_{12}=-100$ in panels (a),  $a_{12}=100$ in panels (b),  $a_{12}=150$  in panels (c), 
and $a_{12}=200$ in panels (d). 
The linear perturbation in the trap pushes the corresponding directly affected component along the direction 
of the perturbation ($x$-direction in our case), modifying the previous density distributions of both elements.
In case of attractive interspecies interaction the effect is just in a redistribution of the overlapped mixture. 
However, when $a_{12}$ is repulsive, and particularly larger than $a_{11}$, the perturbation affects the
spatial distribution in the immiscible phase by changing the previous radial space separation to an axial one,
with the two densities trapped with their maxima at distinct positions.
\begin{figure}[h]
\begin{center}
\includegraphics[width=4.2cm]{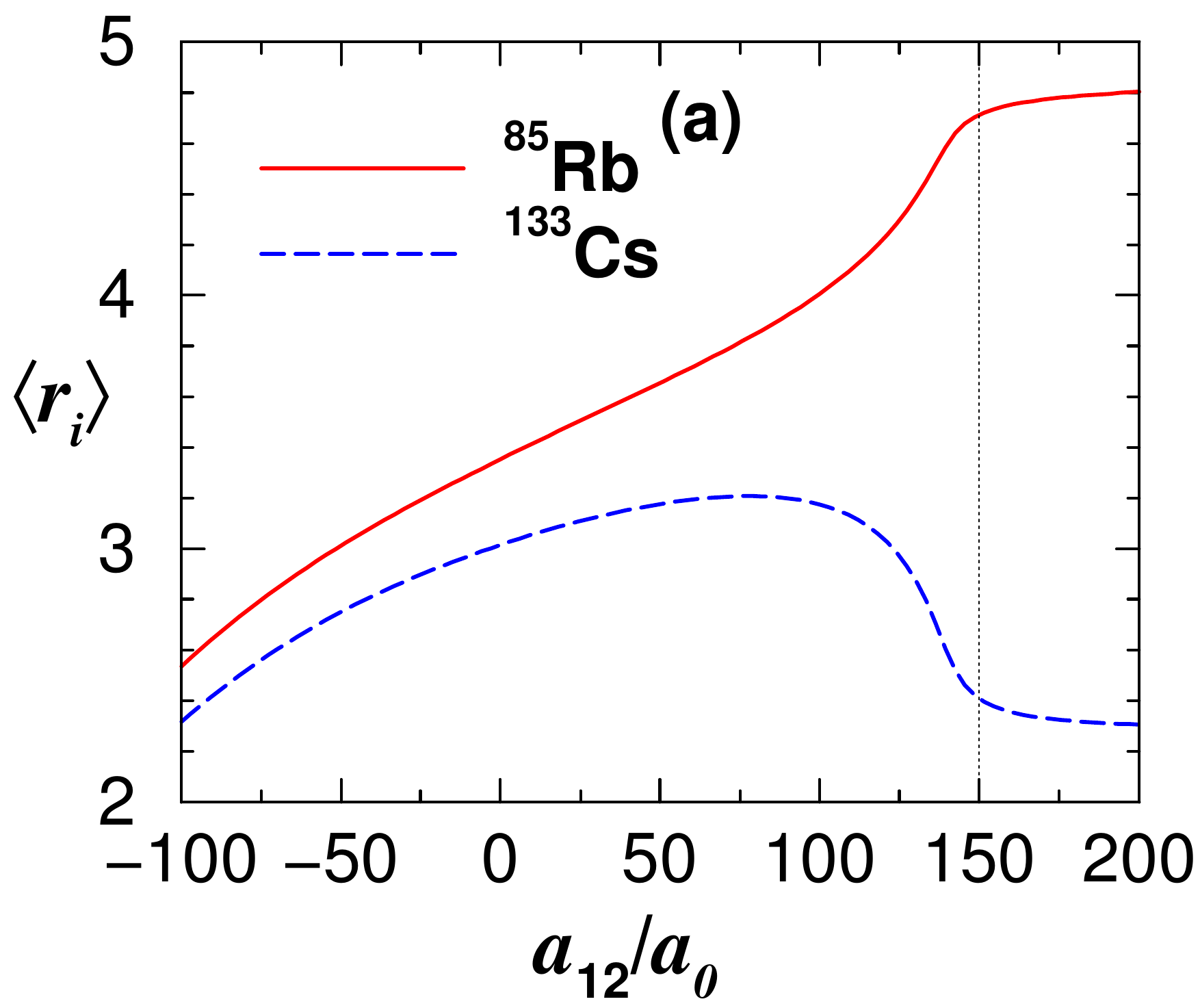}\hspace{-0.7cm}
\includegraphics[width=4.2cm]{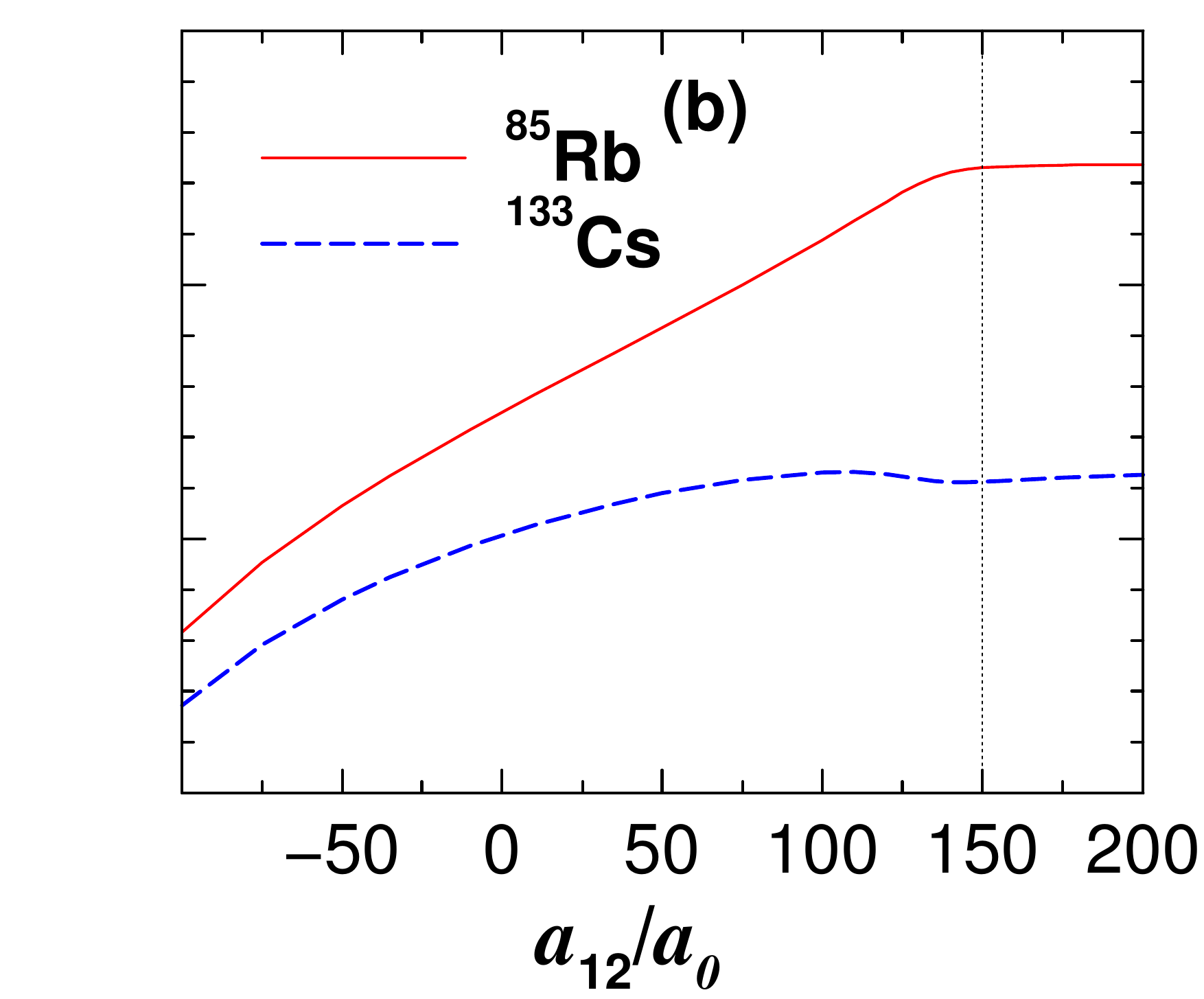}
\vspace{-0.5cm}
\caption{
The root-mean-square radii, $\langle r_i\rangle\equiv \sqrt{\langle x^2+y^2\rangle}$ (in units of $l_{\perp}$), 
for the two components of the non-perturbed [panel (a)] and perturbed [panel (b)] harmonically trapped 
binary mixture, are shown as functions of $a_{12}$ (in units of $a_0$), corresponding to 
Figs.~\ref{f-01} and \ref{f-02} results, respecively, such that $\lambda=10$ and $a_{ii}=150a_0$
The vertical dotted lines (at $\delta=1$) are close to the miscible-immiscible transition.
}
\label{f-03}
\end{center}
\end{figure}

\subsection{Space distribution, miscibility and critical rotation}
In Fig.~\ref{f-03}, by consider both unperturbed [panel (a)] and perturbed [panel (b)] harmonic traps, 
which have been presented in Figs.~\ref{f-01} and \ref{f-02}, we show the behavior of the dimensionless 
root-mean-square (rms) radii $\langle r_i\rangle \equiv \sqrt{\langle r^2\rangle}_i\equiv\sqrt{\langle x^2+y^2\rangle}_i$ 
in terms of the two-body inter-species interaction $a_{12}$, for the two components $i=1,2$ of the binary mixture.
Within our assumption for the intra-species $a_{ii}$,
the corresponding $\delta$ defined by the Eq.(\ref{delta}) is varying from -2/3 till 4/3. 
As verified, the rms radius of the less-massive component (the lighter one, species 1) increases as the interspecies 
interaction increases, in both unperturbed and perturbed situations.
On the other hand, for the unperturbed case, shown in the panel (a) of Fig.~\ref{f-03}, the second component 
(the heavier one, species 2) increases as $\delta$ is  increasing from negative values only till a region where 
$\delta \simeq 2/3$ ($a_{12} \simeq 100 a_0$). For larger values of $a_{12}$, by entering in the more miscible region, 
$\langle r_2\rangle \equiv \sqrt{\langle r^2\rangle}$ starts to reduce due to the radial phase 
separation of the two species, which occurs as the $a_{12}$ becomes more repulsive, saturating to some
small radius at the center of the trap.
By going from the unperturbed to the perturbed regime, the radial distribution is mainly modified at larger repulsive 
values of $a_{12}$, resulting in an axial spatial distribution of the two species. In this space density distribution, the
confined space of component 2 turns out to be larger than before, with the panel (b) of Fig.~\ref{f-03} reflecting the results presented in Fig.~\ref{f-02}.

\begin{figure}[h]
\begin{center}
\includegraphics[width=6.5cm]{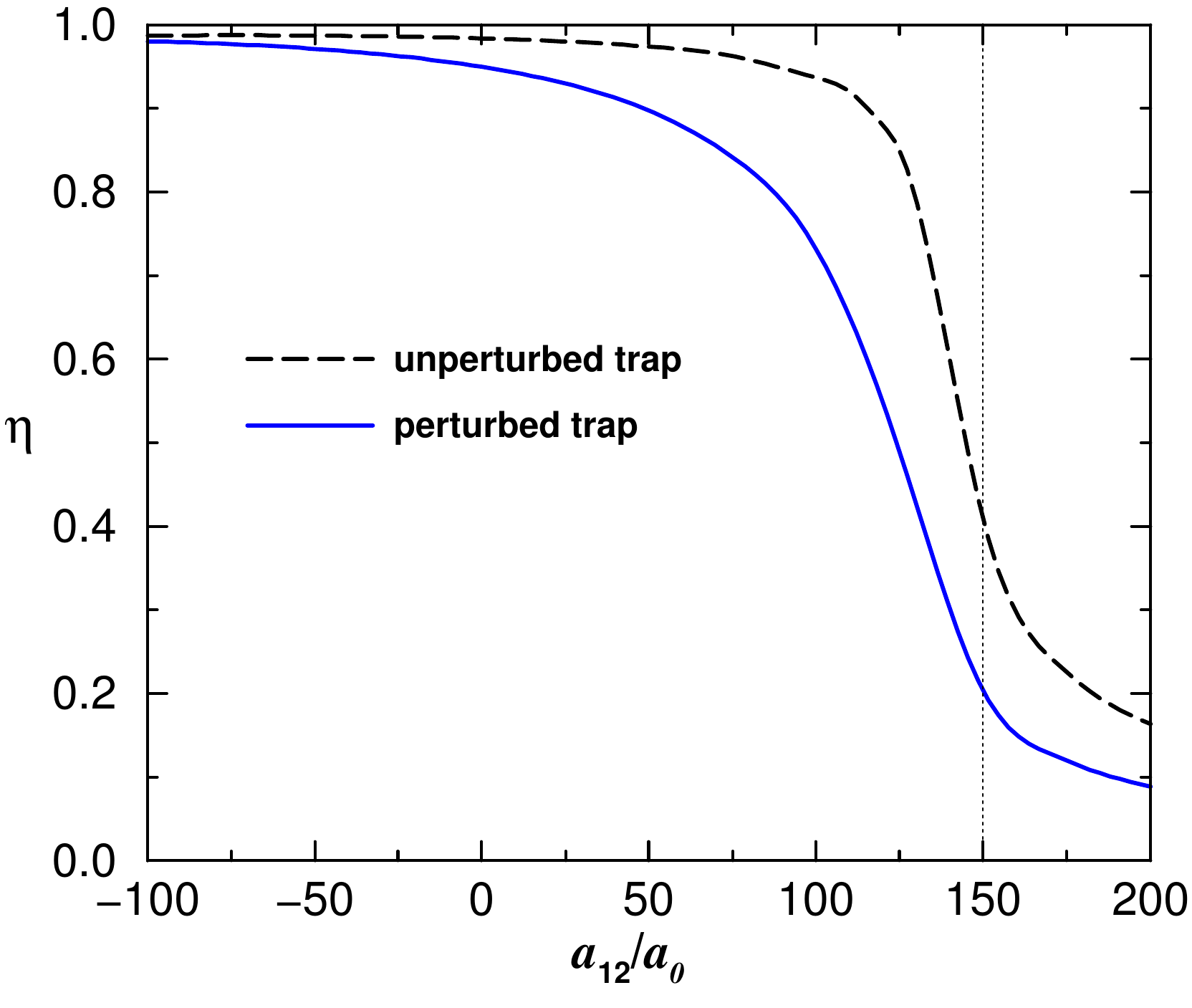}
\vspace{-0.2cm}
\caption{
Miscible-immiscible transition of non-rotating ($\Omega =0$) binary BEC mixture in harmonic non-perturbed (dashed line) 
and linearly perturbed (solid line) traps, in terms of the inter-species $a_{12}$. The parameters are as in Figs.~\ref{f-01}-~\ref{f-03}, 
with the intra-species interactions fixed to $a_{ii}=150a_0.$ The vertical dotted line at $\delta=1$ indicates the miscible-immiscible
transition for equal-mass homogeneous mixture.}
\label{f-04}
\end{center}
\end{figure}

When the system is immiscible (or less miscible), the density distribution of the mixture is no more
homogeneous, with the coupled system being affected by the repulsion between the two species. 
Even for the unperturbed case, we observe that the peak of the density of the lighter element ($^{85}$Rb, 
$i=1$) is deviated from the center of the trap (in case $a_{12}<0$) to a ring 
surrounding the heavier element ($^{133}$Cs, $i=2$) which happens when $a_{12}>a_{11}$), 
as observed from the panels (c) and (d) of Fig.~\ref{f-01}. In this last situation, the component 2 becomes more
strongly confined in the center of the trap due to the inter-species repulsion. 
In the perturbed case that we are considering, we have the density distribution affected by the trap $x-$space 
dislocation of the $^{85}$Rb (component 1), which will also affect the $^{133}$Cs (component 2) density distribution, 
mainly due to the inter-species interaction. So, in both the cases (unperturbed and perturbed), it is relevant to 
verify how the density distribution is affected, by considering the miscibility of the two elements.
The two-component transition from miscible to immiscible is well known as given by the condition expressed in 
Eq.~(\ref{delta})], when considering homogeneous systems~\cite{Chui-1998}. However, for non-homogeneous
 systems is more reliable to consider a parameter that reflects the density distribution of the species, such as the parameter
 $\eta$ that was defined in Ref.~\cite{2017jpco}. By considering this parameter, also given by Eq.~(\ref{eta}),
 a coupled system can be considered as completely miscible if $\eta=1$ and completely immiscible if $\eta=0$, limiting situations
 that can only be approached in both the cases we are considering, as verified in Fig.~\ref{f-04}.
 In this figure, we show $\eta$ as a function of the inter-species scattering length $a_{12}$, for both non-perturbed and 
 perturbed cases. As expected, for non-perturbed harmonic trap, the transition occurs near $\delta=1$, 
 in agreement with Eq.~(\ref{delta}),  with the transition being softened when considering the perturbed trap. 
This miscibility analysis can further explain the density distributions shown in Figs.~\ref{f-01} and \ref{f-02}, in which 
we noticed how the miscibility is affected by the translation in the $x-$position.

The radial distributions of the condensate densities, obtained for $\Omega=0$ and shown in the two panels of
Figs.~\ref{f-03},  are strongly affected by the miscibility of the mixture.  They are relevant to estimate 
the critical frequency  $\Omega_c$ required to generate vortices in a BEC system, as we discuss in the next. 
Even considering that such analysis is limited to homogeneous single component systems, in the present case of non-perturbed 
and perturbed coupled two-component systems it can reflect more directly the critical frequencies in regions where the 
coupled system is more miscible, as for attractive inter-species.
From our calculations, we found appropriate to use a rotation frequency given by $\Omega=0.7$, which is enough larger 
than $\Omega_c$ to help us observe vortex lattice formations in both the cases 
we are discussing in this work.

For the moment, let us recover from Ref.~\cite{1999dalfovo} the discussion related to the full-dimensional 
critical frequency for a single component system with mass $M$. 
By considering a BEC system in axially symmetric trap, with large number $N$ and 
perpendicular Thomas-Fermi (TF) radial distribution $R_\perp$, the critical frequency was analytically derived in 
Ref.~\cite{1997lundh}, being given by 
\begin{equation}
\overline{\Omega}_c = \frac{5}{2}\frac{\hbar}{MR_\perp^2} \ln\left(\frac{0.671 R_\perp}{\xi_0}\right),
\label{OmegaD}
\end{equation} 
where $\xi_0$ is the healing length (also called coherence length), evaluated for the density $n_0$ in the center of the
trap.  For the derivation of (\ref{OmegaD}), it was assumed that the characteristic dimensions of the cloud is large
compared with the coherence length at the center of the cloud, which  is defined by the balance between quantum 
pressure and the interaction energy of the condensate~\cite{1997lundh,1999dalfovo}, in a uniform medium.
The coherence length  can be expressed in terms of the scattering length $a$ and central density $n_{3D}$ 
of the gas without vortex by $\xi=1/\sqrt{8\pi n_{3D}a}$.  
The expression~(\ref{OmegaD}) gives the  {\it full-dimensional} critical frequencies to generate vortices 
in a uniform medium, expected to be valid for each single species independently, when both densities are 
uniformly distributed in a radial space, having the maxima at the same localization $x=y=z=0$. 
When considering a coupled system, the balance between quantum pressure and the interaction energy 
(for each condensate) is being modified due to the miscibility of the components, such that the 
expression (\ref{OmegaD}) is approximately valid only when  the  two species are miscible, not repelling 
each other, limited to regions where $a_{12}\lesssim a_{11}$ (in our case, more precisely, due to the mass
difference, $a_{12} \le 146.3 a_{0}$).
Within our dimensionless units, defining the perpendicular radial distribution aa  
${\cal R}\equiv \left({R_{\perp}}/{l_{\perp}}\right)$, with the {full-dimensional} frequency in terms of 
dimensionless one is $\overline{\Omega}_{c} \equiv {{\Omega}_{c,1} }{\omega_{\perp}}$, 
from (\ref{OmegaD}) we obtain
\begin{equation} 
{\Omega}_{c,1}= 
\frac{5}{2} \frac{1}{{\cal R}^2} 
\ln\left(\frac{0.671 {\cal R}}{{\bar\xi}_{0}}\right) 
\approx\frac{5}{2} \frac{1}{{\cal R}^2} \ln\left({40\,{\cal R}\,|\psi_{01}|
}\right),
\label{Omegac}\end{equation} 
where the dimensionless central healing length is 
${\bar\xi}_{0}\equiv \left[{8\sqrt{\lambda\pi} (a_{11}/l_\perp) N_1 |\psi_{01}|^2}
\right]^{-1/2}$, with the density of the component $1$ at the origin given by 
$n_{0,1}=\sqrt{\lambda/\pi}|\psi_{0,1}|^2$. We are also considering our assumption for the 
intra-species scattering length and number of atoms, 
$N_1=10^4$, $a_{11}/l_\perp=150/(1.89 \times 10^4)$ (such that ${\xi_0}/l_\perp=0.01676 /|\psi_{01}|$).
Within the expectation that Eq.~(\ref{Omegac}) is approximately valid for non-interacting systems, we
can estimate the critical frequency and corresponding radial distribution ${\cal R}$, 
by considering in particular the point where there is no inter-species interaction, $a_{12}=0$.
In this case, the 2D central densities is found $|\psi_{0,1}|^2=$ 0.019, with the critical frequency close to 
0.15, implying that ${\cal R}\approx$ 6.9. This value for the radial distribution is about twice the value of rms radius, 
$\sqrt{\langle (x^2+y^2)\rangle}$, as verified by the component 1 in the panel (a) of Fig.~\ref{f-03}.

\begin{figure}[htbp]
\begin{center}
\includegraphics[width=7.8cm]{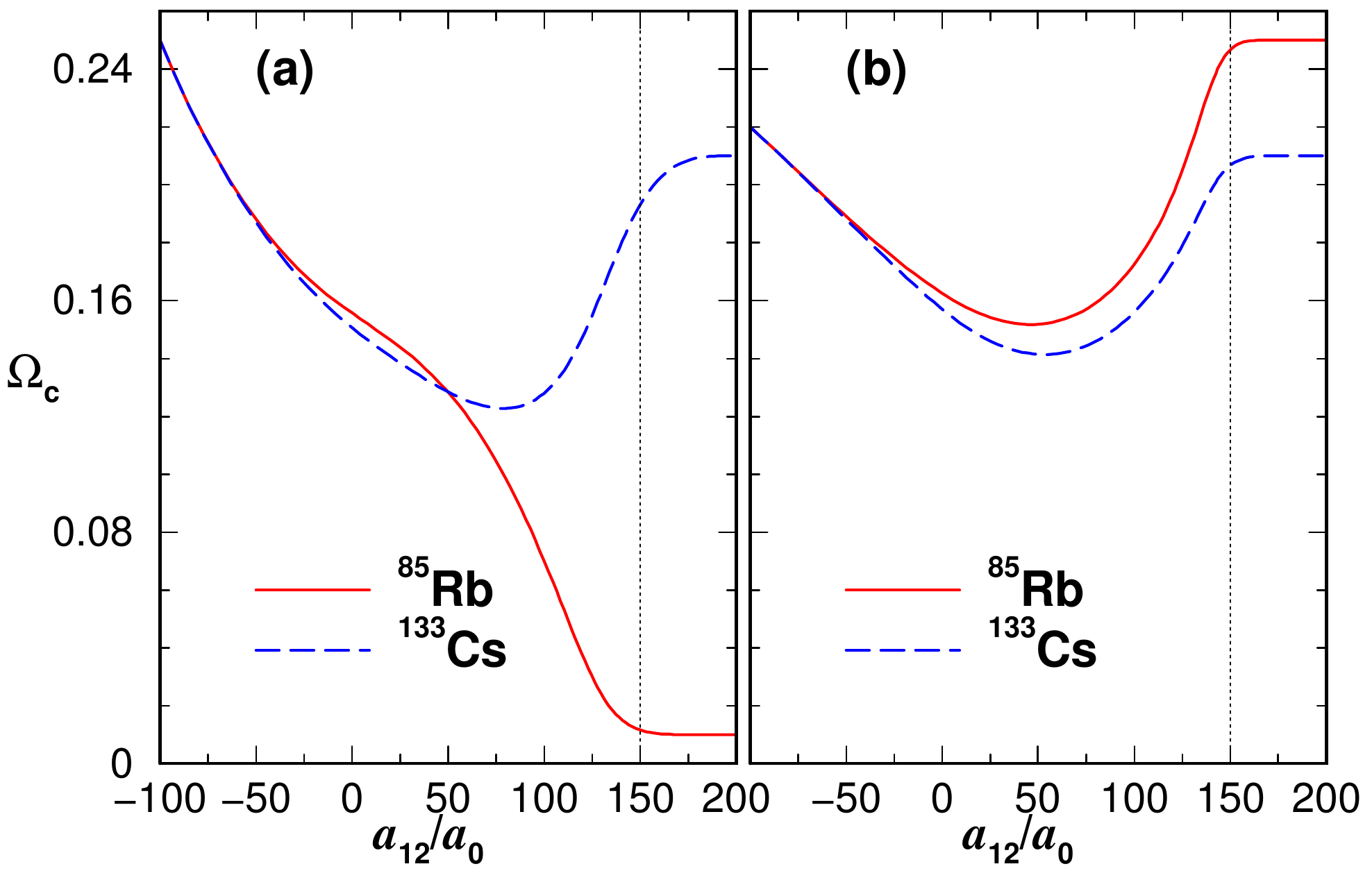}\vspace{-0.2cm}
\caption{
In this figure, the lines (solid-red for $^{85}$Rb, dashed-blue for $^{133}$Cs) are representing the minimum critical frequency 
$\Omega_c$ (as functions of the inter-species $a_{12}$), in order to generate vortices, which can 
exist only above the corresponding curves ($\Omega\ge \Omega_c$).  The non-perturbed and perturbed coupled systems
are, respectively, in the left and right panels. The vertical dotted lines (both panels) are close to the miscibility transition for 
homogeneous systems. These results can be qualitatively 
compared with the ones respectively shown in Fig.~\ref{f-03} for the radii.
In all the cases, $a_{ii}=150 a_0$, with $\Omega$ given in units of $\omega_\perp$.
}
\label{f-05}
\end{center}
\end{figure}

Moreover, about the critical frequencies to generate vortices in a binary coupled systems, it may be worth
mentioning that further theoretical studies are demanding in order to derive an expression more general
then the above (\ref{Omegac}), which can be extended to coupled systems and reflect the computed
results displayed in Fig.~\ref{f-05}.
For such, one can rely in some previous analytical studies (done for single components), in which  the hydrodynamic 
equations have been applied under the TF approximation, as in Refs.~\cite{1997lundh,1999dalfovo},
by considering the two regimes,  with miscible~\cite{2015polo} and 
immiscible~\cite{Chui-1998} phases for the coupled mixture.

In order to calculate the threshold rotation, we should emphasize that it will be helpful the random phase in the 
initial wave function.
Without random phase, a single vortex continues to exist for $\Omega < \Omega_c$.
At the critical rotation frequency ($\Omega_c$) a single vortex enters into the condensate.
 When the rotation frequency is significantly higher than $\Omega_c$, then more vortices will enter into the condensate. 
 The $\Omega_c$ required to create the single vortex in rotating condensates depends on the interaction strength 
 (that determines the radius) and the trap geometry (pancake or cigar). One needs larger $\Omega_c$ for more  
 attractive interaction, which shrinks the condensate radius. In contrast,  the repulsive interaction expands  
 the radius, resulting in smaller $\Omega_c$. 
 Our main results on the critical frequencies are summarized in the Fig.~\ref{f-05}, for both unperturbed [panel (a)] 
 and perturbed [panel (b)] cases.
 By looking the results for the unperturbed case, given in the panel (a), we notice that
 $\Omega_{c}$ of both the first and second component are about the same until $a_{12}\geq 60\,a_0$, when start 
 deviating from each other.  By increasing the inter-species interaction $a_{12}$, the critical frequency of the first 
 component $\Omega_{c_1}$ diminishes due to its radial expansion, which makes more favorable to 
 generate vortices with smaller rotation frequency, saturating at $a_{12}\approx 150\,a_0$  ($\delta=1$). This result
 can be directly associated with the corresponding radial distribution shown in Fig.~\ref{f-03}, in which 
 we do not find big difference in the radius for $a_{12}$ between $150\,a_0$ and $200\,a_0$. 
 In contrast,  $\Omega_{c_2}$ starts to increase from $a_{12}\geq 100\,a_0$, reflecting the respective
 reduction in the radial distribution of the component 2.
 
Next, we consider the $\Omega_c$ for the case of binary mixtures, when the trap of component 1 is linearly 
shifted in the $x-$direction. In this case, the critical frequencies of both components are about the same as the ones
obtained when in the presence of attractive inter-species interaction. 
The repulsive inter-species interaction provides axial phase separation, introducing some small difference between 
$\Omega_{c_1}$ and  $\Omega_{c_2}$ of the mixture.  In this case, from the radial distribution of both systems, 
shown in Fig.~\ref{f-03}, with $\sqrt{\langle r_1^2\rangle}> \sqrt{\langle r_2^2\rangle}$,  one should expect 
$\Omega_{c_1}<\Omega_{c_2}$.  However, we are observing
that $\Omega_{c_1}>\Omega_{c_2}$, which can be explained by the fact that the first component density have its
maximum distribution located a bit far from the center of the trap, as verified from panels (d) of Fig.~\ref{f-02},
 due to the immiscibility of the mixture.

\begin{figure}[h]
\begin{center}
\includegraphics[width=0.49\linewidth]{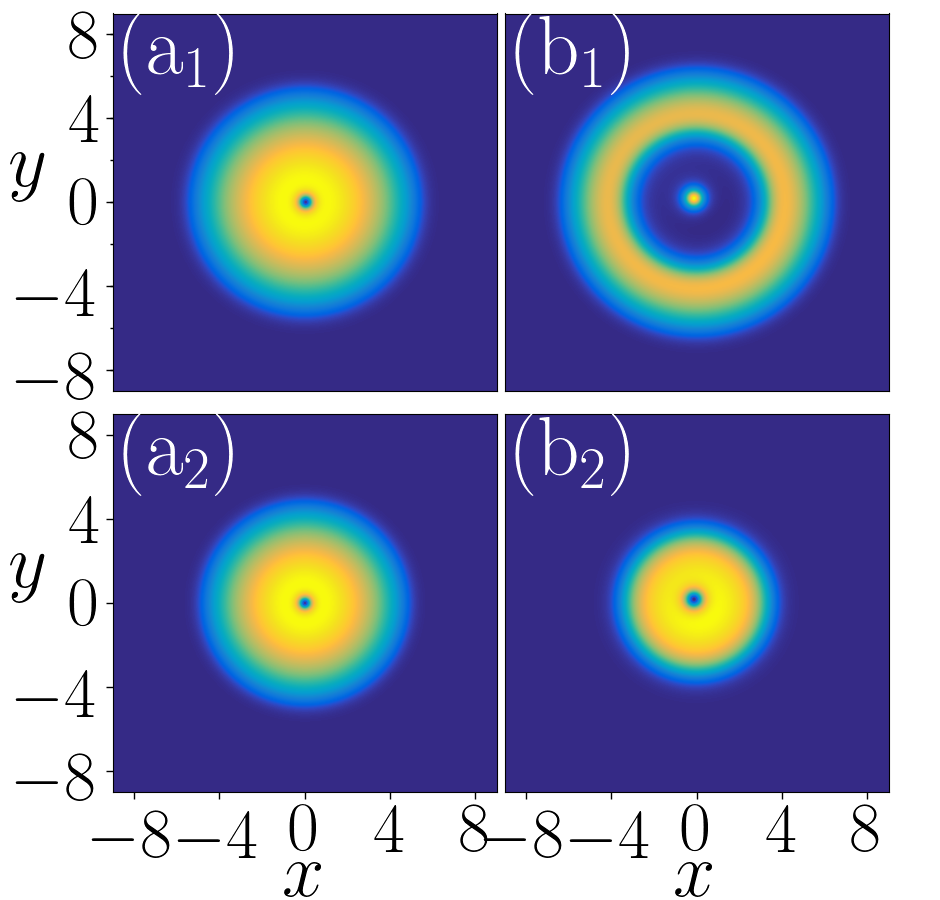}
\includegraphics[width=0.49\linewidth]{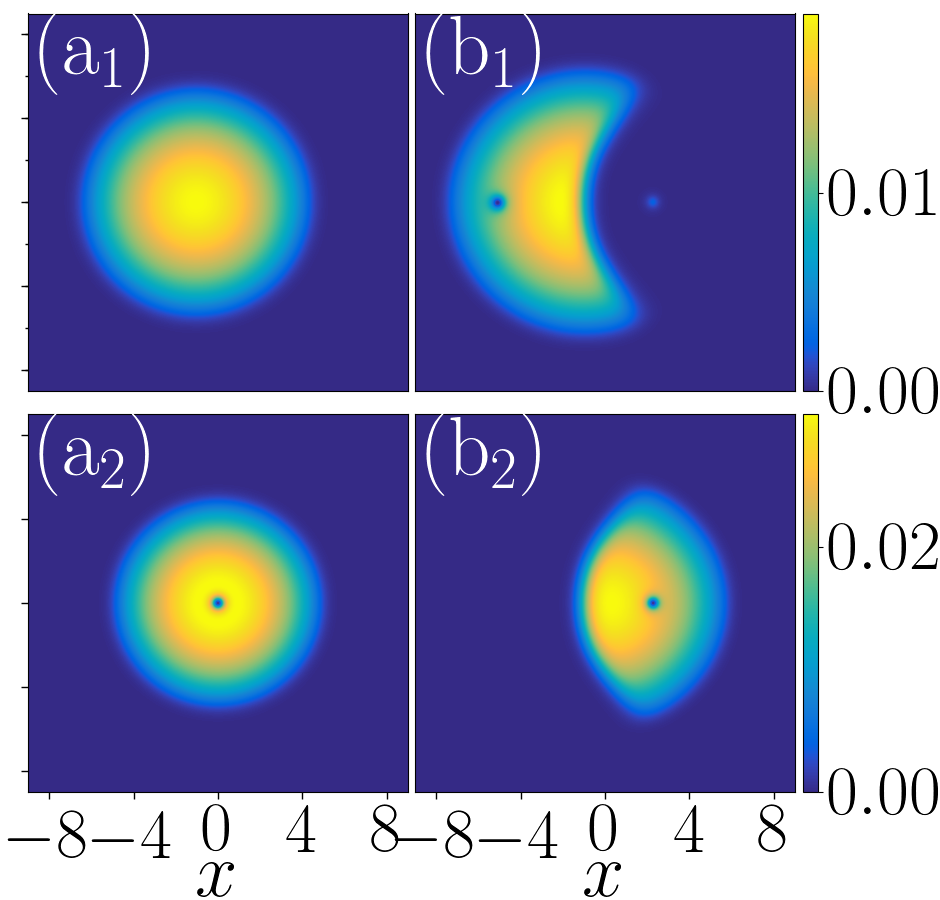}
\includegraphics[width=0.49\linewidth]{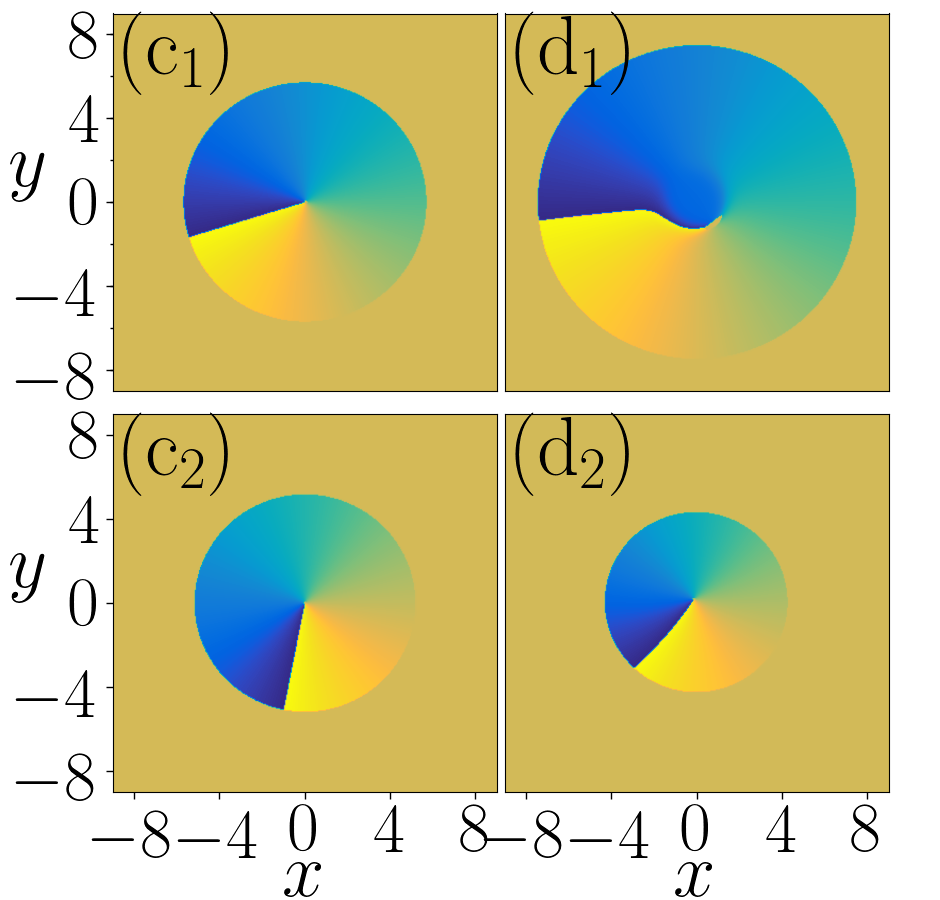}
\includegraphics[width=0.49\linewidth]{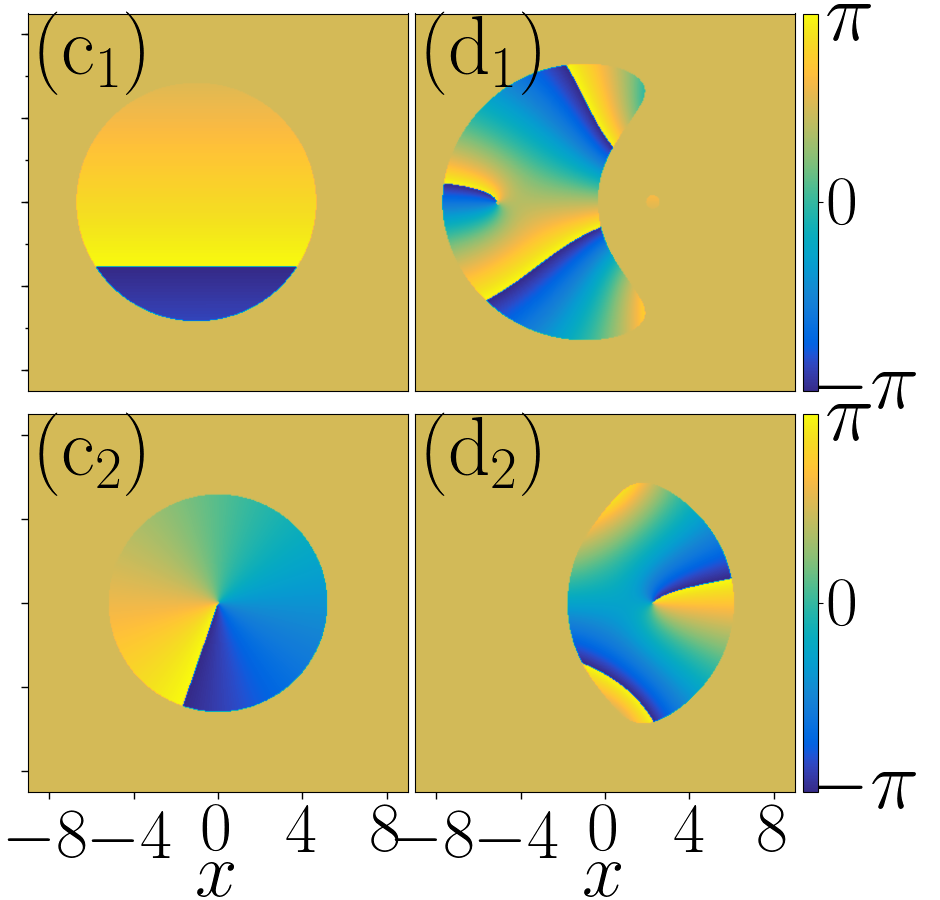}
\vspace{-0.2cm}
\caption{
Sample results of 2D densities $|\psi_{i=1,2}(x,y)|^2$ (units $l_\perp^{-2}$) in the ($x,y$) plane
(space unit $l_\perp$) for the coupled non-perturbed (left upper set of panels) and linear perturbed 
(right upper set of panels) systems.
Correspondingly, the phase diagrams are shown in the next two lower rows of panels (c$_i$) and (d$_i$). 
In the non-perturbed cases (left set), the panels (a$_i$) and (c$_i$) are for $a_{12}=0$ ($\delta=0$) with $\Omega=0.17$, 
with panels (b$_i$) and (d$_i$)  for $a_{12}=150 a_0$ ($\delta=1$) with $\Omega=0.22$. 
In the perturbed case (right set), the panels (a$_i$) and (c$_i$) are for $a_{12}=0$ ($\delta=0$) with $\Omega=0.16$,
with panels (b$_i$) and (d$_i$)  for $a_{12}=150 a_0$ ($\delta=1$) with $\Omega=0.25$.  
In all the cases, $a_{ii}=150 a_0$, with trap aspect ratio $\lambda=10$.
}
\label{f-06}
\end{center}
\end{figure}

In order to verify how the vortices are being generated when we are close to the critical limit of rotation, we 
present some results in Fig.~\ref{f-06}, by considering, respectively, both the unperturbed and perturbed trap cases. 
For these sample results, the densities for the species $i=1,2$ are represented in the upper part of the figure by 
two sets with four panels [ (a$_i$) and  (b$_i$)], in which the left set is for the unperturbed case, with the right 
set referring to the perturbed case. Correspondingly, we have the respective phases shown in two sets with four 
panels [ (c$_i$) and  (d$_i$)] in the lower part of the figure.
For the non-perturbed case, shown in the left sets of Fig.~\ref{f-06} (densities and phases),  we are considering 
the non-interacting case $a_{12}=0$ in the panels (a$_i$) and (c$_i$), with rotation frequency $\Omega=0.17$; 
whereas in the panels (b$_i$) and (d$_i$) we have $a_{12}=150a_0$ ($\delta=1$) with $\Omega=0.22$.
In particular, for the case with $\delta=1$, in which from Fig.~\ref{f-05} we have $\Omega_c\sim 0.01$, one can verify 
the occurrence of vortices in the low-density region of component 1[panel (b$_1$)] from the corresponding phase 
diagram, given in the  panel (d$_1$).
Analogously, for the perturbed case shown in the right sets of Fig.~\ref{f-06} (densities and phases), 
we have $a_{12}=0$ with $\Omega=0.16$ in the panels (a$_i$) and (c$_i$); with the repulsive case, 
$a_{12}=150a_0$ ($\delta=1$), with $\Omega=0.25$, being shown in the panels (a$_i$) and (c$_i$).
As shown in both the cases represented in the Fig.~\ref{f-06}, with $a_{12}=150a_0$, we have already 
entered in the immiscible regime of the mixtures. Radial separation
of the mixture is verified in case we have non-perturbed trap system, with the lighter element ($^{85}$Rb) surrounding the
heavier element ($^{133}$Cs), which is in the center. These results follows in correspondence with the ones verified 
in Figs.~\ref{f-01} and \ref{f-03}(a), when $\Omega=0$, for the relative position of the element densities in the trap.

\begin{figure}[htbp]
\begin{center}
\includegraphics[width=8cm]{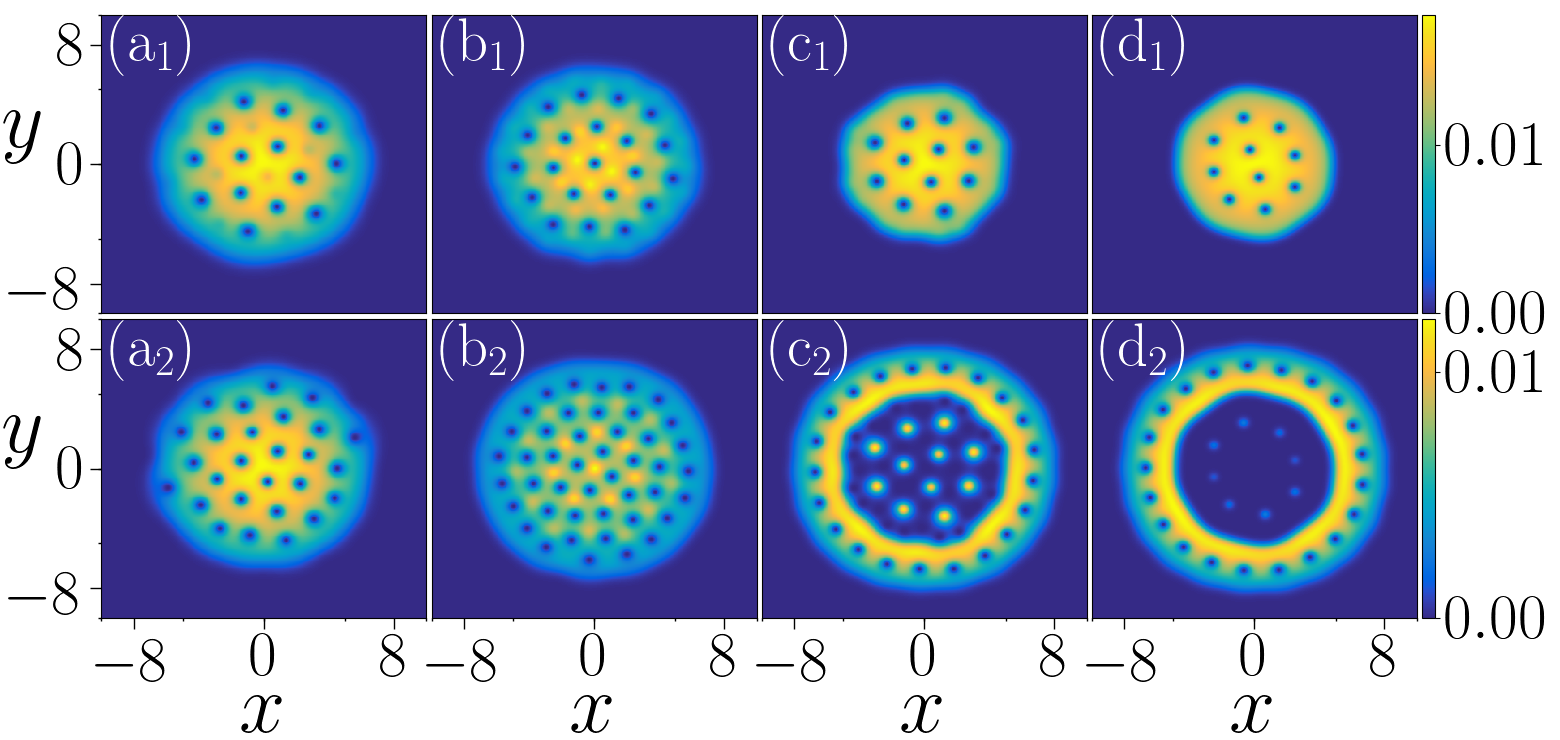}\vspace{-0.2cm}
\caption{
Vortex-lattice structures in the 2D coupled densities, $|\psi_i(x,y)|^2$ (units $l_\perp^{-2}$), for the binary mixture 
$^{85}$Rb$-^{133}$Cs, using 
pancake-like harmonic trap with aspect ratio $\lambda=10$ and rotation frequency $\Omega=0.7\omega_\perp$, 
in the $(x,y)$ plane (space unit $l_\perp$).  With $a_{ii}=150a_0$, the $a_{12}$ goes from attractive
$a_{12}=-50\,a_0$ [panels (a$_i$)], providing complete overlap of the densities; to three repulsive cases, with 
$a_{12}=50\,a_0$ [panels (b$_i$)], $150\,a_0$ [panels (c$_i$)] and $200\,a_0$ [panels (d$_i$)], in 
which (c$_i$) and (d$_i$) are in almost complete immiscible radial separation, with the lighter element 
occupying the center of the trap.
}
\label{f-07}
\end{center}
\end{figure}

\begin{figure}[h]
\begin{center}
\includegraphics[width=8cm]{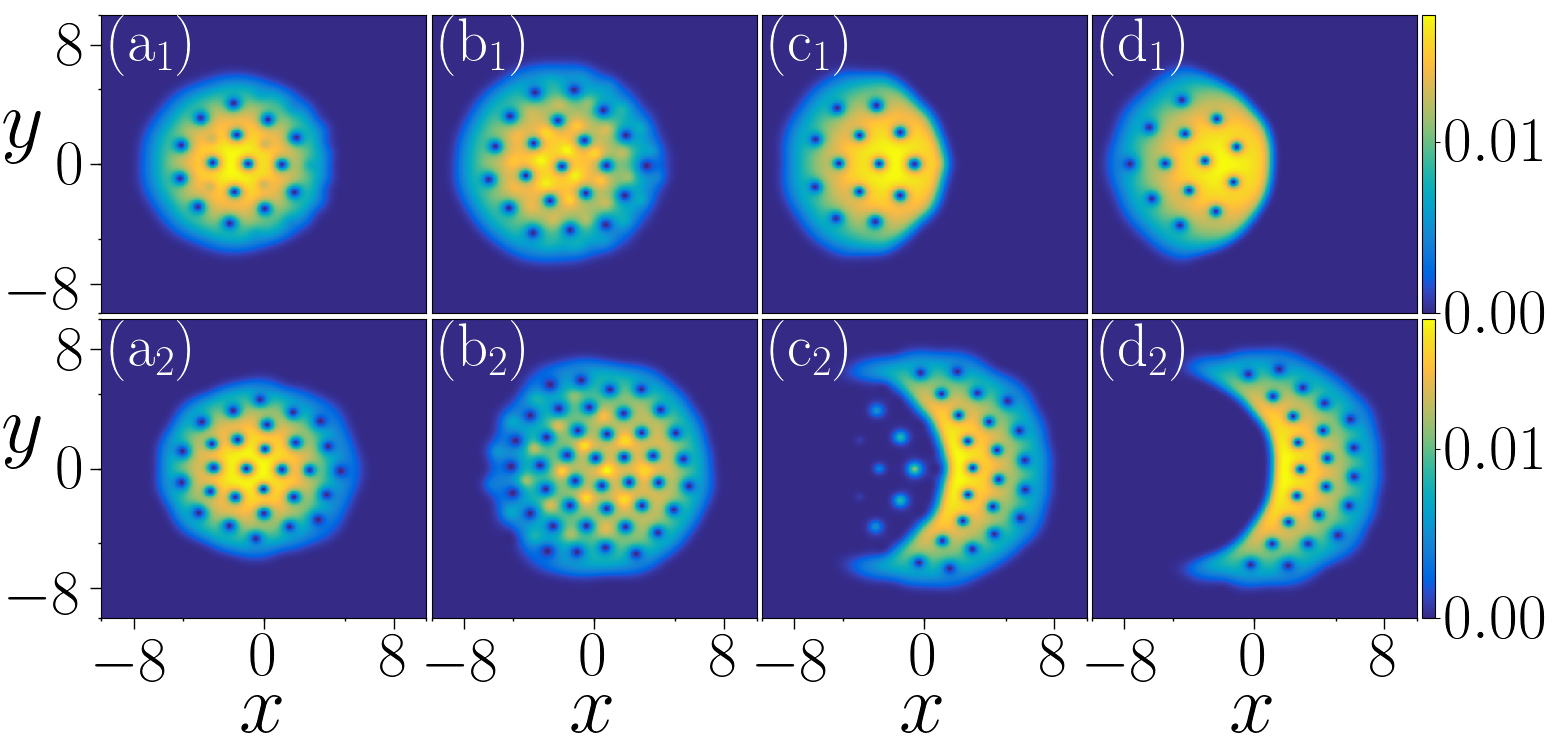}\vspace{-0.3cm}
\caption{
This figure shows the vortex-lattice structures corresponding to the results shown in Fig.~\ref{f-07},
when the harmonic trap of component 1 is perturbed in the $x-$direction ($\nu_1=1, \nu_2=0$).
Except for the perturbation, the other parameters are as in Fig.~\ref{f-07}, 
with $\Omega=0.7\omega_\perp$, $\lambda=10$,  and with $a_{12}=-50\,a_0$ [panels (a$_i$)], 
$a_{12}=50\,a_0$ [panels (b$_i$)], $150\,a_0$ [panels (c$_i$)] and $200\,a_0$ [panels (d$_i$)]. 
The almost complete immiscible condition (c$_i$) and (d$_i$) are providing space separation 
with both species having their maxima side-by-side in the $x-$direction.  
}
\label{f-08}
\end{center}
\end{figure}

Next, we consider a larger and fixed rotation, given by $\Omega=0.7$, which has shown to be enough to create some 
lattice patterns of vortices in both components of the mixture. These results are shown in Fig.~\ref{f-07}, for the unperturbed 
case; and in Fig.~\ref{f-08} for the perturbed case. In both the cases, we assume four values for the two-body interaction, going from 
an attractive cases, $a_{12}=-50a_0$ ($\delta=-1/3$), to three repulsive ones, with  
$a_{12}=50a_0$ ($\delta=1/3$), $a_{12}=150a_0$ ($\delta=1$), and $a_{12}=200a_0$ ($\delta=4/3$). As been clearly verified
in both figures, the cases with $\delta\ge 1$ shown in the panels (c$_i$) and (d$_i$) correspond to the immiscible cases.
Here, even for the unperturbed case, we can already observe a striking result due to the mass-imbalance sensibility 
of the mixture which is emerging as the rotation frequency is increased.  As observed from Fig.~\ref{f-01}, where
we have $\Omega=0$ and from upper set of Fig.~\ref{f-06} where the rotation is quite small, 
in particular for the case with $\delta\ge 1$, 
we have radial space separation, the coupled element interchange their spatial separated positions for higher frequencies, 
such that with $\Omega=0.7$ we can already verify the lighter element occupying the center of the trap, 
with the heavier element surrounding the first (in opposite locations, as compared with the case of low or zero 
rotation frequency).   
Due to the immiscibility and confinement to a smaller radius, the number of vortices of 
the component 1, which is in the center, becomes smaller than the number of vortices appearing in the component 2.  

For the perturbed case,  shown in Fig.~\ref{f-08}, the $x-$dislocation of the component 1 density affects the 
radial-shaped structure separation of the mixture, such that the coupled system turns out 
to display an axial phase-space separation with the species 1 shifted from the center. Due to the fact that component 1
is located in a smaller radius, this translation in the $x-$position will also affect the component 2, which was surrounding
the first before the perturbation, as verified in particular for the immiscible regime shown by the panels (c$_i$) and 
(d$_i$).  
In conclusion, the Figs.~\ref{f-07} and \ref{f-08} are presenting our main results concerning the mass-imbalance 
sensibility of the mixture when considering the relative space separation of the two components, particularly for 
repulsive inter-species interaction larger than the intra-species ones. These results provide further support to the 
conclusions obtained in Ref.~\cite{2019-Kishor} about the relevance of the mass difference in a binary rotating mixture. 

\section{Summary and Discussion}\label{sec4} 
 We have studied the $^{85}$Rb-$^{133}$Cs  binary BEC mixture in rotating non-perturbed and perturbed harmonic traps. 
 First, we examine the equilibrium non-rotating ground states by changing the inter-species contact interaction strength. 
 Due to the mass-imbalance of the mixture, we observe a radial phase separation between the two confined species 
 when the inter-species interaction $a_{12}$ is repulsive and larger than the corresponding intra-species  $a_{11}=a_{22}$, 
 with both species confined in unperturbed harmonic traps. 
 The mean-square radii of the binary BEC mixtures are obtained by verifying that $\langle r_1^2 \rangle$ of the first 
 component (the lighter one) increases as $a_{12}$ becomes more repulsive, with $\langle r_2^2 \rangle$ (of the
 heavier element) reverting its increasing behavior as $a_{12}$ approaches the value of $a_{ii}$, in agreement
 with the mass-dependent condition (\ref{delta}) for the miscible-immiscible transition of homogeneous mixtures. 
 This behavior occurs due to the strong inter-species repulsion. Next, to study the rotational properties of the 
  $^{85}$Rb-$^{133}$Cs BEC mixture, we introduce a non-zero angular momentum rotation frequency 
  $\Omega$ (in units of the transversal trap frequency $\omega_{1,\perp}$), 
 estimating the corresponding critical values $\Omega_{c_i}$ to observe a single vortex in the binary mixture. 
 In agreement with the radial distributions given by the  rms radii, the $\Omega_{c_1}$ continuously decreases as 
 $a_{12}$ becomes more repulsive, till reaching a minimum frequency value for $a_{12}> a_{ii}$. 
 However, in the case of the heavier species, as $a_{12}$ increases, 
the initial decreasing behavior of $\Omega_{c_2}$ reaches a minimum in correspondence with the radial behavior.
After that, $\Omega_{c_2}$ starts increasing till a saturating limit for $a_{12}> a_{ii}$. 
This tendency that occurs for the non-perturbed coupled system is consistent with the immiscibility of the coupled 
mixture, such that the two species start to repel each other strongly for $\delta>1$. 
Without a perturbation in the trap, the distribution of the two densities
becomes radially separated in the immiscible regime, with the heavier element located in the inner part, such that larger
value of $\Omega$ is needed to start vortex generation.
A linear perturbation applied to the element 1 will affect this radial distribution, which happens in the immiscible regime,
having both element distributions with their maxima at separate positions, implying in having their
density with similar side-by-side distributions. So, in both the cases, $\Omega_{c_i}$ decreases in the miscible region
as $a_{12}$ increases, till some independent minima which occur when the system starts becoming immiscible,
saturating at some value $a_{12}> a_{ii}$.

 When the coupled system is confined by unperturbed harmonic traps, in this immiscible regime we have the less massive 
 species surrounding the more massive one, for the non-rotating case $\Omega=0$ or when the rotation is not large enough. 
 As we increase the frequency of the rotation, a transition occurs resulting in an interchange on the radial space 
distribution of the mixture.  As shown, in the immiscible regime, for $\Omega=0.7$ one can already verify that the 
peak of the density of the more massive element is no more in the center, but outside the center of the trap, 
in a ring surrounding the lighter one, which has moved to the center. 
The less massive element, in this case, becomes confined in the inner center region, under the pressure of the
more massive element. 
This is a relevant ``centrifugal effect" that, in some way,  could be expected, as been relevant in order to 
calibrate experimental realizations with rotating mass-imbalanced BEC mixtures, such as $^{85}$Rb-$^{133}$Cs.  
The effect can be further explored for different mass-imbalanced systems, by considering the corresponding 
frequencies for the transitions.

By studying the effect of linear perturbation (along the $x-$direction) in the trap of one of the mass-imbalanced components 
(which we choose the lighter one of the $^{85}$Rb-$^{133}$Cs mixture), we are providing some more elements 
for a possible simple experimental realization in cold-atom laboratories, with rotating mass-imbalanced binary non-dipolar
systems. 
The presence of linear perturbation in the trap pushes the first component along that direction, altering the 
density distribution and the corresponding associated vortices in the rotating regime. 
In the attractive case and in the miscible regime  ($a_{12}<a_{ii}$), we just observe some expected changes 
in the vortex patterns, already explored in several other works. However, it is in the more repulsive case 
(immiscible regime) that we have verified the more significant changes, with the previous radial space separation being 
changed to a side-by-side phase separation between the elements of the mixture. 
Within such side-by-side spatial separation in the immiscible
regime ($\delta>1$), both densities have their trapped regions basically governed by their mass differences.
For both elements, the vortex-pattern structures are distributed within the corresponding available spatial 
regions. 

\section*{Acknowledgements}
The authors acknowledge partial support received from Conselho Nacional de Desenvolvimento 
Cient\'\i fico e Tecnol\'ogico (CNPq) [Procs. 304469/2019-0 (LT), 153522/2018-6 (RKK) and 
306920/2018-2 (AG)], 
Funda\c c\~ao de Amparo \`a Pesquisa do Estado de S\~ao Paulo (FAPESP) 
[Contracts 2016/14120-6 (LT), 2016/17612-7 (AG)]. RKK also acknowledge support from
Marsden Fund (Contract UOO1726).

\end{document}